\documentclass[12pt]{article}

\usepackage{amsmath}
\usepackage{amssymb}

\newtheorem{theorem}{Theorem}[section]

\newtheorem{definition}{Definition}[section]

\numberwithin{equation}{section}

\usepackage{amsmath}
\usepackage{amssymb}

\begin{document}
	\begin{center}
		{\bf Thoughts on Quantum Events, 
		Quantizing Gravity 
		\\
		and von Neumann 
		Algebras}
		\\
		\vskip 0.2cm
		Stephen Bruce Sontz
		\\
		Centro de Investigaci\'on en Matem\'aticas, A.C. 
		(CIMAT)
		\\
		Guanajuato, M\'exico

	\end{center}
 
\begin{abstract}
This short document presents no new results.
It is a collection of related ideas that 
may lead to a way of quantizing gravity. 
To achieve this it may be relevant to use
my recent new approach to
quantum theory, which 
is based on {\em quantum events}. 
Given that most standard quantum texts do not 
define a mathematical object which corresponds
to a {\em physical event}, the definition of a 
quantum event is motivated and discussed 
in detail. 
This in turn leads to the idea that 
the mathematical object that contains 
enough quantum events 
for describing a 
quantum system is a von Neumann algebra. 
The next idea is 
that the kinematics of special relativity 
(as embodied in flat Minkowski space) 
should be quantized as a first step in 
the program of the 
quantization
of general relativity, that is, 
quantizing Einstein's equation. 
It must be emphasized that the events
of special relativity are viewed as part
of the kinematics, rather than of the dynamics
of that theory. The same holds for general
relativity. 
Similarly, I am viewing quantum events as part
of the kinematics of quantum theory, even 
though some find this to be 
counterintuitive. 

An Appendix shows how ``collapse'' is 
easily understood in this new approach 
to quantum theory.

\end{abstract}

\section{Introduction}
It is not known whether it is possible to 
find a quantum theory (QT) that in some 
reasonable sense is a quantization of 
general relativity (GR). 
As a minimum the 
quantum theory should have some sort of limit
as Planck's constant $ \hbar \to 0 $ that 
has some agreement with GR.  
But I claim that a clearer understanding 
of QT in itself is required 
in order to construct a self-consistent 
quantization. 
Unfortunately, QT in general is presently 
plagued with some perplexing problems whose 
proposed solutions remain controversial. 
Among these problems are finding 
the correct roles in QT of the following: 
the state (or the wave function), 
the measurement process, entanglement of 
distant, seemingly independent systems,     
the probabilistic collapse of the state 
versus  
the deterministic evolution of the state 
given by the Schr\"odinger equation. 
In the recent paper \cite{new-approach}
and in more detail in a recent book \cite{new-approach-book}
I have presented 
a new approach to quantum theory, showing 
that these 
are not intractable problems, but only confusions that disappear in a new clearer 
approach to quantum theory that is based on 
a probability theory of 
quantum events 
that is appropriate for
quantum theory. 
As such it merits the name
{\em quantum probability theory}. 

Also another problem is to 
specify what the probabilities of QT apply to in nature. 
The approach in \cite{new-approach} and \cite{new-approach-book} is 
that probabilities refer to what can possibly happen and,
more particularly, with what frequency. 
That which can happen is usually called an
{\em event} in common parlance. 
Consequently a basic feature of QT is to 
define a mathematical structure which 
corresponds to a physically occurring event.  
Furthermore, any probability theory 
should be a calculus for describing the 
frequency of events, and most particularly, 
sequences of multiple events.
A probability theory that only assigns a  
probability of an individual, isolated 
event\footnote{Born's Rule 
in standard quantum theory 
refers to only one measurement, which is a type of event. 
This limitation inherent in Born's Rule leads to the
infamous quantum ``paradoxes'' 
and its ``weirdness''.}
is not relevant, since science is based on the
correlations of multiple events. 
There is even a folkloric saying to this effect:
``All probabilities are conditional
probabilities.''

So this note begins with an exposition of 
the mathematical structure corresponding to
a physical event. 
While this is not a new concept, going back 
at least to 1981 
(see \cite{beltrametti}), it seems not
to be well known in the physics community. 
So a detailed discussion is given to justify 
the mathematical definition of a 
{\em quantum event} 
or in short an {\em event}. 
This will be a mathematical object that corresponds 
to, but is distinct from, a physical event.  

Then it is maintained that the encompassing 
mathematical 
scenario for QT has to accommodate all 
events relevant to a given quantum system. 
It turns out that this will be a von 
Neumann algebra. 

With this understanding of QT in place, we 
turn to the problem of quantizing GR. 
However, GR is also a theory based on 
events, 
since the points is spacetime
have coordinates that label the ``when and
where'' of physically occurring events. 
This parallel between the basic mathematical
structures of QT and GR is then used to propose 
a possible way that GR could be quantized. 
A simpler initial problem would be one with no
dynamics, that is, with only kinematics 
playing a role in QT and GR. 
Explicitly, I am proposing the quantization 
of SR as a 
possible starting point for the quantization of GR.

\section{Events}

A basic organizing idea of 
\cite{new-approach} and \cite{new-approach-book}  
is that there are certain 
mathematical structures which correspond to 
physically occurring events. 
These abstract structures are called
{\em quantum events} or, more simply, 
{\em events}.
This sort of abuse of terminology is already 
present in standard quantum theory, where
self-adjoint operators are called 
{\em observables}, since they correspond to 
physically occurring observables. 
However, 
the idea that events play a central role in 
quantum theory - or even they they appear
in that theory - does not seem to be widely 
known. 
So some discussion of what this mathematical 
structure is, and why it is so named, is
in order. 

In the first place, physical events are 
physical observables. 
Thus it seems 
reasonable to suppose that the 
corresponding mathematical structure $ E $ 
should be a self-adjoint operator acting in 
some Hilbert space $ \mathcal{H} $. 
So we will assume that $ E = E^{*} $. 
Also, a physical 
event either occurs or it does not. 
Therefore, the spectrum of
$ E $ should consist of 
(at most, as we shall soon see) two 
eigenvalues, 
each corresponding to one of these 
possibilities.
Traditionally, when there are two options, 
these are labeled (with no loss of generality
or intuition) 
by the numbers $ 0 $ and $ 1 $. 
Here the convention will be that the eigenvalue 
 $ 1 $ corresponds to the occurrence of the 
 physical event and that the eigenvalue $ 0 $
corresponds to its not occurring. 
So we will assume that the spectrum of $ E $
satisfies 
$ \mathrm{Spec}\, E \subset \{ 0, 1 \} $. 
Notice that the self-adjoint operator $ E $ 
has a bounded spectrum and therefore is a
bounded operator whose domain is the Hilbert 
space $ \mathcal{H} $. 
Symbolically, we have 
$ E : \mathcal{H} \to \mathcal{H} $. 
Moreover, we have the decomposition of 
$ \mathcal{H} = \mathcal{H}_{0} \oplus 
\mathcal{H}_{1}  $, where 
$$
\mathcal{H}_{\lambda} := \{
\psi \in \mathcal{H} \,|\, E \psi = \lambda \psi \}
$$
is the spectral subspace associated with 
the eigenvalue $ \lambda \in \{ 0 , 1 \}$. 
Since $ E^{2} = E $ holds on each subspace 
$ \mathcal{H}_{\lambda} $, it also holds on 
$ \mathcal{H} $. 
This discussion motivates the following.  
\begin{definition}
Let $  \mathcal{H} $ be a Hilbert space. 
A {\rm (quantum) event} is a linear, bounded 
operator 
$ E : \mathcal{H} \to \mathcal{H} $ that
satisfies $ E = E^{*} = E^{2} $. 
\end{definition}

Note that conversely any event $ E $ has 
$ \mathrm{Spec}\, E \subset \{ 0, 1 \} $ 
and 
$ \mathcal{H} = \mathcal{H}_{0} \oplus 
\mathcal{H}_{1}  $, as described above. 
Since the zero operator $ 0 $ and the identity
operator $ I $ are events, we see that both
$ \mathrm{Spec}\, E = \{ 0 \} $ and 
$ \mathrm{Spec}\, E = \{ 1 \} $ are possible. 
However, in all other cases, 
$ \mathrm{Spec}\, E = \{ 0, 1 \} $.  

It turns out that this such an operator is 
well known in functional analysis and has 
another name: {\em projection}. 
So one may well ask what this change of 
nomenclature\footnote{Other names used are: 
proposition, yes/no experiment, test, question, filter.
I do not know who first used the term 'event' in 
this context, though \cite{von-neumann} is a possibility.} 
is all about. 
After all, any naming convention is as good as any other. 
As Hilbert pointed out, the mathematical 
structure of geometry would not be changed if 
the three words ``point, line, plane'' were replaced 
by any other sequence of three (distinct!)
words. 
Of course, we use ``point, line, plane'' 
because it aids intuition. 

So what further justification 
is there, beyond the comments above, 
to use the 
word ``event'' when the word ``projection'' 
is well established? 
Note that the answer can only 
be based on its value 
as an aid to intuition, since 
the exact terminology does not matter 
mathematically. 
And here we are concerned with what might be 
called ``physical intuition'' rather than 
``mathematical intuition''. 
Actually, ``projection'' does aid 
mathematical (or even better geometrical) 
intuition. 
So the justification must have something to 
do with the correspondence of these 
mathematical objects with physically occurring
events. 

One way to justify this terminology is by 
using the spectral theory for 
possibly unbounded, self-adjoint 
operators acting in a Hilbert space. 
This theory in its general form is not 
widely known in the physics community, 
and its precise role in physics is not widely 
known in the mathematics community. 
Consequently I will present some results
(but not any proofs, which can be found in 
the literature, say \cite{weidmann}) 
of the general theory. 
But since this is rather abstract, I will 
later present a specific case that should be
more familiar. 

So here is an outline of the general 
theory for any (even unbounded) self-adjoint 
operator $ T $ acting in a Hilbert space 
$ \mathcal{H} \ne 0 $. 
It then follows that the spectrum of $ T $, 
denoted $ \mathrm{Spec} \, T $ and consisting 
of all the ``generalized eigenvalues'' of $ T $, 
is a non-empty closed subset of the real
numbers $ \mathbb{R} $. 
A nice fact is that $ \mathrm{Spec} \, T $ 
is a bounded subset of $ \mathbb{R} $ if and 
only if $ T $ is a bounded operator. 
It is known that $ \mathrm{Spec} \, T $ 
is an important object associated with $ T $. 
Indeed, thinking of $ T $ as corresponding to 
a physical observable, then the actual values 
observed are exactly the numbers in 
$ \mathrm{Spec} \, T $, neither more nor 
less.\footnote{Of course, these is an idealization, 
since measurements are not exactly precise.} 
Unfortunately, $ \mathrm{Spec} \, T $ is not 
sufficient for completely understanding $ T $. 
However, spectral theory provides another 
mathematical structure that turns out to 
be equivalent to $ T $. 
Here is its definition. 
\begin{definition}
A {\em projection valued measure (pvm)} on the 
real line $ \mathbb{R} $ with values in a 
Hilbert space $ \mathcal{H} $ is a mapping 
$ B \mapsto E (B) $ for every 
Borel \footnote{A Borel set is open or 
closed or any other set more or less constructible 
from these. Explicitly, it is the $\sigma$-algebra generated by
the open sets.} 
set $ B \subset \mathbb{R} $ with these properties: 
\begin{itemize}
	
	\item $ E (B) $ is a projection operator
	acting on  $ \mathcal{H} $. 
	
	\item 
	$ E (\emptyset) = 0 $, the zero operator, 
	and 
	$ E (\mathbb{R}) = I $, the identity 
	operator. 
	(Here $ \emptyset $, the empty set, and 
	$ \mathbb{R} $ are Borel sets. Also, 
	$ 0 $ and $ I $ are projections.) 
	
	\item 
	If $ \{ B_{j} \} $ is a countable family of
	pairwise disjoint Borel subsets of 
	$ \mathbb{R} $ (meaning 
	$ B_{j} \cap B_{k} = \emptyset$ 
	if $ j \ne k $), then we have 
	$ E (\cup_{j} B_{j}) = \sum_{j} E (B_{j}) $. 
    Here the possibly infinite sum of operators 
    on the right side is meant in the sense that
    for every $ \psi \in \mathcal{H} $ we have
    that
    $\psi = \sum_{j} E (B_{j}) \psi $ with 
	convergence of this sum being in the 
	usual norm topology of $  \mathcal{H} $. 
	
\end{itemize}

\end{definition}

Not surprisingly, a pvm being a measure has an
associated integral.
This enters in the following. 
\begin{theorem}
{\bf (Spectral Theorem)}
Suppose $ T $ is a possibly unbounded self-adjoint 
operator acting in a Hilbert 
space $ \mathcal{H} $. 
Then there exists a unique pvm $ E_{T} $ 
such that 
$$
T=\int_{\mathbb{R}} \lambda\, E_{T} (d \lambda). 
$$
Conversely, any pvm $ E $ defines a self-adjoint 
operator $ T $ by the formula 
$$
T := \int_{\mathbb{R}} \lambda\, E (d \lambda). 
$$
Moreover, this gives a one-to-one and onto
correspondence between the set of all self-adjoint operators acting in 
$ \mathcal{H} $ and the set of all pvm's on 
the real line $ \mathbb{R} $ with values
in $ \mathcal{H} $. 

\end{theorem}

The exact definition of the integral 
of a function with respect to a pvm is given in 
standard texts of functional analysis, 
such as \cite{weidmann} 
where the proof of the Spectral 
Theorem can also be found. 
As we shall see later, this theorem 
generalizes - in
some sense - that $ T $ is diagonalizable. 

But for now, I want to explain in terms of physics what a pvm is. 
As the notation already suggests, a pvm 
can also be called a quantum event valued 
measure.\footnote{Actually, I used this terminology in the introductory text 
\cite{my-qm-book}.} 
Let's consider $ E_{T} (B) $ for some Borel
set $ B \subset \mathbb{R} $ and some 
self-adjoint $ T $. 
What is this event physically? 
It turns out that to answer this question
we will appeal to another aspect of spectral 
theory. 

\begin{theorem}
\label{spec-map-thm}
{\bf (Spectral Mapping Theorem)} 
If $ f : \mathbb{R} \to \mathbb{R} $ is a 
Borel function\footnote{This includes characteristic functions of a Borel set and all 
continuous functions.} which is bounded 
and $ T $ is a 
self-adjoint operator, then 
$$
 f (T):= \int_{\mathbb{R}} f (\lambda) 
     \,    E_{T} (d \lambda)
$$
is a bounded, self-adjoint operator satisfying
$$
\mathrm{Spec} \, f(T) = f ( \mathrm{Spec} \, T), 
$$
where $ f ( \mathrm{Spec} \, T):= 
\{ f (\lambda) \,|\, \lambda \in \mathrm{Spec} \, T\} $ 
is simply the image under $ f $ of 
the spectrum of $ T $. 
\end{theorem}

Physically, if $ T $ corresponds to a physical 
observable, whose observed values are the
numbers in $  \mathrm{Spec} \, T $, then
$ f (T) $ corresponds to the physical 
observable whose values are the images under 
the action of $ f $ on the values of $ T $. 
For example, any device measuring the values of $ T $ can be connected to a computer which 
changes the values $ \lambda $ given by the device to the values $ f(\lambda) $, that is,
the computer is programmed to compute the
values of $ f $. 
Then the resulting combination is a new device that 
measures the values of $ f (T) $. 

Next, we do a calculation for the case 
$ f = \chi_{B}  $, the characteristic 
function for a Borel set 
$ B \subset \mathbb{R} $. 
Here we define 
\begin{equation*}
\chi_{B} (\lambda):= \Big\{ 
\begin{array}[2]{cc}
1 & \mathrm{for~} \lambda \in B,
\\
0 & \mathrm{for~} \lambda \notin B. 
\end{array}
\end{equation*}
So we have 
$$
\chi_{B} (T)= \int_{\mathbb{R}} \chi_{B} (\lambda) \, E_{T} (d \lambda) = 
\int_{B} \, E_{T} (d \lambda) = 
E_{T} (B). 
$$
The last two equalities come from the integration theory of the pvm.
As the reader may note they are
analogous to equalities in standard 
measure theory.  

Next, 
$ \mathrm{Spec} \, E_{T} (B) = \mathrm{Spec} \, \big(\chi_{B} (T)\big) = 
\chi _{B} (\mathrm{Spec} \, T) 
\subset \{ 0, 1 \}$ by Theorem 
\ref{spec-map-thm}. 
So, $ 1 \in \mathrm{Spec} \, E_{T} (B) $ 
corresponds the existence of 
some $ \lambda \in \mathrm{Spec} \, T $ 
such that $ \chi_{B} (\lambda) = 1 $ 
which in turn is equivalent to 
$ \lambda \in B \cap \mathrm{Spec} \, T $. 
Intuitively, this says $ E_{T} (B) $ 
is the event which occurs if and only if 
``$ T $ has a value in $ B $''. 
To emphasize this intuition we introduce 
the notation 
$$
    T \in B := E_{T} (B).
$$
Notice that the three symbols on the left 
hand side 
only make sense in this combination. 
Of course, a self-adjoint operator is not 
an element in a Borel set. 

It is at this point that probability enters 
quantum theory for a given self-adjoint operator 
$ T $ acting in $ \mathcal{H} $. 
To do this we take $ \psi \in \mathcal{H} $ 
with $ || \psi || = 1 $. 
Then for every Borel set 
$ B \subset \mathbb{R} $ we define
$$
   \mu_{\psi} (B):= 
   \langle \psi, E_{T} (B) \psi \rangle =
   || E_{T} (B) \psi ||^{2} \in [0, 1]. 
$$

It follows that $ \mu_{\psi}  $
is a probability measure on $ \mathbb{R} $
as given in the usual presentation of 
Kolmogorov of probability theory as a part 
of measure theory. 
Informally, one often reads this as ``the 
probability that $ T $ takes a value in $ B $
given the state $ \psi $''. 
Of course, $ \psi $ itself is not a state, but 
rather represents the set 
$ \{ e^{i \theta} \psi \,|\, \theta \in  
\mathbb{R} \} $, which is the state. 
The technical, though trivial, detail is that 
the probability measure $ \mu_{\psi} $ is
unchanged if $ \psi $ is replaced by 
$ e^{i \theta} \psi $ for any $ \theta \in  
\mathbb{R} $. 

Now we can go one step beyond Kolmogorov 
probability by defining 
$$
P_{\psi} (E) := \langle \psi , E \psi \rangle 
=
|| E \psi ||^{2} \in [0, 1], 
$$
where  
$ \psi \in \mathcal{H} $ 
with $ || \psi || = 1 $ 
and $ E $ is an event acting on 
$ \mathcal{H} $. 
Informally, one can read this as ``the 
probability that the event $ E $ occurs 
given the state $ \psi $''. 
Notice that this probability of an event 
has nothing to do with Borel sets or
the $ \sigma $-algebras of measure theory.  

Now let's see how this general spectral theory 
plays out in a special case. 
We suppose that $ T = T^{*} $ has a spectrum 
consisting only of eigenvalues, i.e., 
$$
\mathrm{spec \, T} = 
\{ \lambda_{j} \}_{j=1}^{N}
$$
for distinct eigenvalues, that is, 
$ \lambda_{j} \ne \lambda_{k} $ 
for $ j \ne k $. 
Here $ N $ can be finite or $ N = \infty $. 
For each eigenvalue $ \lambda_{j} $ we define 
its associated eigenspace by 
$$
V_{j} := \{ \psi \in \mathcal{H} \,|\, 
T \psi = \lambda_{j} \psi \},
$$
which we assume to be finite dimensional
and hence closed. 
Because $ T $ is self-adjoint, 
these subspaces are pairwise orthogonal.
It follows that $ \mathcal{H} $ has an 
orthonormal basis consisting only of 
eigenvectors, namely, each basis element 
belongs to some $ V_{j} $.
We define $ E_{\lambda_{j}} $ to be the 
projection onto the closed subspace $ V_{j} $. 
Since these subspaces are pairwise orthogonal, 
we have 
$ E_{\lambda_{j}} E_{\lambda_{k}} = 0 $ 
for $ j \ne k $. 
Since $ T $ acts on $ V_{j} $ by multiplication 
by $ \lambda_{j} $, which is how 
$ \lambda_{j} E_{\lambda_{j}} $ also acts
on $ V_{j} $, it becomes apparent that
$$
   T = \sum_{j} \lambda_{j} E_{\lambda_{j}},
$$
which is the statement of the spectral 
theorem in this case. 
It is also called the {\em diagonalization} 
of $ T $, 
since the matrix of $ T $ (in the orthonormal 
basis defined above) has the 
eigenvalues appearing along the diagonal, each 
a finite number of times, and all entries 
off the diagonal being zero. 
Also notice that we have a resolution of the
identity operator $ I $ 
$$
    I =  \sum_{j} E_{\lambda_{j}}
$$
in the sense that for all 
$ \psi \in \mathcal{H} $
we have that 
$$
\psi =  \sum_{j} E_{\lambda_{j}} \psi 
$$
where the possibly infinite sum on the right 
converges in the norm topology of 
$ \mathcal{H} $. 

Given all of this, we can now define the pvm 
$ E_{T} $ associated to $ T $ for a Borel
set $ B \subset \mathbb{R} $ by  
$$
E_{T} (B) := 
\sum_{\{ j \,|\, \lambda_{j} \in B \}}
E_{\lambda_{j}}. 
$$
Consequently, $ E_{T} (B) $ is the event 
corresponding to $ T $ 
having the value of some eigenvalue 
$ \lambda_{j} $ which is in $ B $. 

Notice  that this discussion so far, with the 
exception of the formulas for probabilities, 
has not used states. 
Also notice that all the events of 
this theory are 
quantum events (associated to a given 
Hilbert space, of course). 
There are no ``classical'' events or 
``classical measuring devices'' mentioned. 
But by putting the emphasis on the role of
quantum events, this leads to a deeper 
understanding of the role of projections 
and serves to further justify calling them 
events. 
This has to do with a shortcoming in most 
texts on quantum theory, including mine 
\cite{my-qm-book}. 
The point is that in those presentations only 
the formula for a single event, usually one of 
the form $ T \in B $, is given. 
Sometimes $ T \in B $ is read as ``$ T $ is 
measured to have a value in $ B $''. 
This leads to the idea that measurements play a 
special role in quantum theory. 
But they do not in \cite{new-approach}
and \cite{new-approach-book}, where 
measurements just correspond to events of the
form $ T \in (a,b) $, where $ (a,b) $
is a ``small'' non-empty open 
interval \footnote{
Definition: $ (a,b):= \{ \lambda \in 
\mathbb{R} \,|\, a < \lambda < b \} $} 
in $ \mathbb{R} $. 
Without formulas for the probability of multiple 
consecutive events (including the important case 
of conditional probability), 
the standard approach lacks relevant tools. 
It is no small wonder that the so-called 
``paradoxes'' of quantum theory pertain to the 
analysis of multiple events. 
In particular, the ``collapse'' of the state and 
entanglement are difficult to explain. 
But the approach of \cite{new-approach} and \cite{new-approach-book}
gives explicit formulas for the probabilities 
of multiple events. 
The calculations based on these formulas give a 
unified way for dealing with these ``paradoxes''. 
Moreover, being formulas which give explicit 
numerical results, one can check those results 
with experiments and observations. 

For example, ``collapse'' of the state turns out 
to be one step in a two step algorithm for 
computing {\em one} conditional probability of 
a sequence of two events. 
(See Appendix A.) 
That is, ``collapse'' is not something that 
happens in a quantum system; it is not a
quantum event. 
L\"uders Rule is similarly understood. 
Also, the explicit example of entanglement 
of two 
electrons in a spin singlet state is 
calculated  in \cite[pp. 21--23]{new-approach}
in various ways using conditional 
probability 
with results that 
agree with experiment. 
In short, entanglement is a consequence of 
simple formulas for the probabilities of 
multiple events. 
Only events are entangled at a fundamental level 
in this approach. 
For example, entangled 
states \footnote{These are defined in tensor 
product Hilbert spaces. Entanglement is not
limited to this special situation, but is 
defined as lack of independence of events. See \cite{new-approach}.} 
 are understood in 
terms of related tangled events. 
There also are no anthropomorphic ``classical'' 
observers (aka Alice and Bob) in this approach. 

All in all, the success of the approach in 
\cite{new-approach} in resolving
``paradoxes''
by using a computational 
analysis in 
terms of events, is another reason for 
renaming projections as (quantum) events.

\section{Quantization}

Quantization is the construction of a 
mathematical model adequate for the 
description a quantum system. 
It does not necessarily have anything to do 
with a 
mathematical model of a ``corresponding'' 
classical system, even though that is 
a possibility. 
For a quantum system we have seen that such 
a model should include enough events to 
describe what actually can physically occur 
and possibly nothing more. 
Moreover, a reasonable collection of states, 
possibly including density matrices, should be 
available in the model in order to be able to 
calculate all the relevant probabilities of 
the quantum system. 
To start off, it is well known what is the 
mathematical structure associated to a Hilbert 
space that also accommodates a large collection 
of events. 
This is a von Neumann algebra.\footnote{To be
adequate for quantum theory, this algebra should be non-commutative.}

Often in quantization schemes there is given 
only a Hilbert space $ \mathcal{H} $, and the 
implicit assumption is that the von Neumann 
algebra is the type I factor 
$$
\mathcal{L} ( \mathcal{H} ):= 
\{  T :  \mathcal{H} \to  \mathcal{H} \,|\, T \mathrm{~is~linear~and~bounded} \}. 
$$ 
This is often adequate, but sometimes it is not. 
It is known that other types of von Neumann 
algebras arise in quantum statistical mechanics 
and in quantum field theory. 
The use of $ \mathcal{L} ( \mathcal{H} ) $
when that is not appropriate has been called 
the Type I error. 

So  the first step in quantization should be 
to find the appropriate von Neumann algebra 
for the quantum system being considered. 
This thereby introduces the set of events  
of that von Neumann algebra as the relevant 
set of events for the quantum system. 
While a von Neumann algebra $ \mathcal{V} $ 
can be considered as an abstract mathematical 
object, for the present purposes of this note 
I wish to consider only those explicitly given 
as 
$\mathcal{V} \subset \mathcal{L}(\mathcal{H})$ 
for some Hilbert space $ \mathcal{H} $. 
Consequently, the choice of $ \mathcal{H} $ 
is also a part of the quantization, but 
it is not enough in and of itself. 

In the next step an appropriate set of states 
(including density matrices) must be chosen. 
These will be used to calculate 
all the probabilities 
associated to the sequences of multiple 
events of the quantization. 

So far, the discussion of this note concerns 
only the kinematics of a quantum system. 
How a quantum system changes in time 
(that is to say, its dynamics) 
is a further consideration of no immediate
pertinence for now. 
(See \cite{new-approach} for details.) 

It is important to note that the kinematics 
described so far is not sufficient for doing 
all of quantum theory. 
For example, there has been no mention of spin. 
Nonetheless, even this minimal treatment of 
quantum theory is sufficient for understanding 
``collapse'' and entanglement. 
The measurement problem is also easy to resolve.

\section{Gravity}
The current knowledge in physics 
has not yet sufficed to 
decide whether a reasonable quantum theory of 
gravity exists. 
This is usually taken to mean an improvement, 
via a quantization,  
of GR with no changes to QT. 
However, some maintain that to reconcile these two theories, it is QT which must 
be modified. 
This remains an open problem, but I will 
explore the possibility that gravity can be 
quantized while leaving quantum theory 
unchanged. 
So the 
rest of this note will be a discussion of how 
one might implement an approach for 
quantizing gravity. 
This is clearly a daunting task.  

However, one way to begin to think about such 
a putative quantization is to consider the 
purely kinematical case first. 
We have already seen the basic kinematical 
outline of QT 
in terms of events and their 
probabilities.  
A simple kinematical approach to GR
is to consider just the flat 
Minkowski spacetime $ \mathbb{R}^{4} $ with 
no dynamics. 
Then the question is how does one find a 
quantization of this spacetime. 
Of course, spacetime itself is a set 
of events (together with more structure to 
be sure). 
So the idea is to quantize this set of events 
as a family of non-commutative 
von Neumann algebras 
$ \mathcal{V}_{\hbar} $, depending on Planck's 
constant $ \hbar > 0 $, in such a way that the 
event structure of $ \mathcal{V}_{\hbar} $ 
tends to the event structure of  
$ \mathbb{R}^{4} $ in the limit as 
$ \hbar \to 0 $. 
Of course, the exact type of limit to be used 
here is part of the problem of finding this 
quantization. 

Let me mention that the event structure of 
the Minkowski space $ \mathbb{R}^{4} $ 
is sometimes criticized as being 
idealized, since each of its points is a
mathematical event, which is meant to model 
a physical event,
and points are rather singular objects. 
This is not a fatal flaw with 
$ \mathbb{R}^{4} $. 
Rather, 
one can instead consider the relevant events 
of this spacetime to be all the non-empty open 
subsets of $ \mathbb{R}^{4} $. 
Then the points themselves in $ \mathbb{R}^{4} $ 
are no longer the principal interest. 
For example, the events could be all those in 
the commutative von Neumann algebra  
$ L^{\infty} ( \mathbb{R}^{4} ) $, that is, 
all the characteristic functions of Borel 
subsets of 
$ \mathbb{R}^{4} $ modulo function equal 
almost everywhere. 
This, of course, removes the points of 
$ \mathbb{R}^{4} $ from the set of events,
contrary to their original role as the basic
events of SR. 

This is as far as this quantization program 
goes for now. 
Clearly, it is only an idea that requires much 
further development to be acceptable as a 
quantization of Minkowski space. 
Besides this initial incomplete kinematical 
theory, one would have to continue to the 
dynamical theory and show how the dynamics 
of GR, namely Einstein's equation, 
is realized as an 
appropriate limit of the dynamics of quantum 
theory. 
While this is a formidable project, it seems 
to be quite different from other approaches 
to this situation and so may merit more 
attention.  

The idea that SR - with no dynamics - should be
quantized before doing GR is contrary to the idea
that one should always have a Hamiltonian $H$ when
doing QT. 
But actually, no dynamics is the special case $H = 0$. 
This indeed should be the easiest case. 

But there is something here that seems to be rather 
counterintuitive for many people. 
And that is that events are being considered as part
of the kinematics and not of the dynamics. 
Now this is completely standard in SR, while 
I have found that this aspect of my approach to 
QT is very disconcerting for some colleagues. 
And it gets worse in GR, where the very structure
of the totality of events (known as spacetime) is
itself subject to dynamics. 
Notice that the very word `spacetime' conceals the fact
that its points are events.  
The usual formalism in physics uses a dichotomy,
one could even say a dualism, that phrases physical
theories solely in terms of kinematics and
dynamics. 
This distinction has pre-Socratic roots in the ideas
about what exists (and therefore
is subject to possible change)
and about what is change (and therefore to what 
possible existent entities it applies).   
Heraclitus held the extreme view that nothing 
exists in and of itself, but only change is real.
Parmenides took the totally contrary view that 
nothing changes and there is only static 
existence. 
Other ancients, most notably Aristotle, tried to
reconcile existence and change.  
It is interesting that
scientists and philosophers continue to deal with 
these ideas, trying to sort all that out to this day. 
My point is that these issues could well be critical 
for understanding the relation of QT and GR.

\section{Von Neumann Algebras}

The remarks so far suggest the importance 
of von Neumann algebras. 
It seems that in practice quantization is 
done with a particular type of 
von Neumann algebra, namely a factor, which 
is characterized as having center 
isomorphic to $ \mathbb{C} $. 
Of course, the literature is vast and so 
I might well be wrong. 
There does not appear to be any reason to 
restrict quantization to just the case of
factors. 
But, if the center is non-trivial, what 
could that mean? 
One possibility is that the center, taken 
by itself, represents a classical system or, 
more likely, a small deformation 
of a classical theory. 
This would then be another way of understanding 
what a quantization of a classical system is. 

Moreover the relevant von Neumann algebras of
QT are non-commutative $ C^{*} $-algebras 
and may be amenable 
to methods of non-commutative geometry. 

\vskip 0.2cm 
This note ends with these vague speculations.

\section{Appendix A} 
Here is how ``collapse'' is handled in this 
approach. 
We start with events $ E $ and $ F $, 
acting in a Hilbert space $ \mathcal{H} $, 
such that 
$ E $ can occur first and $ F $ later. 
Then given 
$\psi \in \mathcal{H}$ with $ ||\psi||=1$, the 
{\em consecutive probability} 
for $ E $ occurring 
first and then later $ F $ is defined as
$$
   P_{\psi} (E,F):= || F E \psi ||^{2} \in 
   [0,1]. 
$$
Notice that $ FE $ is itself an event if and only 
if $ E F = F E $.  
In that particular case
$$
 P_{\psi} (E,F) =  P_{\psi} (EF) = 
  P_{\psi} (FE)
$$
is the probability of that single 
event  $ E F = F E $. 
However, when $ E $ and $ F $ do 
not commute, this 
consecutive probability is not consistent with 
standard Kolmogorov probability theory 
based on measure theory. 
It is something else. 
While this has led some to say that this is 
not a ``real probability'', it turns out that 
E.P.~Wigner was not one of those. 
Indeed, this consecutive probability is 
called Wigner's Rule. 
(See \cite{wigner}.)
Using Wigner's Rule as a starting point, we can 
define the {\em conditional probability}
of $ F $ occurring, given that $ E $ occurred 
earlier (for a given $ \psi $ as above) as: 
$$
P_{\psi} (F \,|\, E ):= \Big\{ 
\begin{array}[2]{cl}
 P_{\psi} (E,F)/ P_{\psi} (E)  & 
\mathrm{if~}  P_{\psi} (E) \ne 0,
\\
0  & \mathrm{otherwise.}
\end{array} 
$$
Note that this is analogous to the definition 
in standard probability theory. 
Moreover, in the case that 
$ P_{\psi} (E) \ne 0 $ we have 
\begin{equation}
\label{conditional-as-collapse}
P_{\psi} (F \,|\, E ) = 
\dfrac{P_{\psi} (E,F)}{P_{\psi} (E)} \!=\!
\dfrac{|| FE \psi ||^{2}}{|| E \psi ||^{2}} 
\!=\! 
\big|\big| F  \dfrac{E \psi}{|| E \psi ||} \big|\big|^{2} =
|| F \tilde{\psi} ||^{2} =
P_{\tilde{\psi}} (F), 
\end{equation} 
where $ \tilde{\psi} := E \psi/|| E \psi || $ 
is a unit vector in $  \mathcal{H} $. 

In the language of the standard texts, such as 
\cite{my-qm-book}, 
the transformation $ \psi \to \tilde{\psi} $ 
is called the ``collapse'' of 
the state (or of the 
wave function) which is induced (or even caused) by the
first event $ E $. 
Note that the formula for 
$ \tilde{\psi} $ depends only 
$ \psi $ and $ E $. 
Given this situation it is claimed that the 
probability of a subsequent occurrence of $ F $ 
is given by $ P_{\tilde{\psi}} (F) $, which is
the probability of the single event $ F $, 
given the ``collapsed'' state\footnote{Actually, the state determined by $ \tilde{\psi} $,}
$ \tilde{\psi} $. 
However, in this language, which is  
also known as Born's Rule and 
is restricted to
only statements of probabilities of single 
events, 
it is neither evident nor ever stated that 
$ P_{\tilde{\psi}} (F) $ is the conditional 
probability $ P_{\psi} (F \,|\, E ) $, 
which is the content of 
\eqref{conditional-as-collapse}. 
Of course, the only events here are 
$ E $ and $ F $. 
The change $ \psi \to \tilde{\psi} $ is not 
an event, but rather one step in a two step 
algorithm for computing the {\em one} 
conditional probability 
$ P_{\psi} (F \,|\, E ) $. 
(The second step is the computation of 
 $ P_{\tilde{\psi}} (F) $, which looks like 
Born's Rule but is not.) 
 
 A three step algorithm for computing this 
 one conditional probability is given in the 
 second expression in \eqref{conditional-as-collapse}. 
 That algorithm is as follows:
 \begin{enumerate}
 \item Compute the numerator $ P_{\psi} (E,F) $. 
 
 \item Compute the denominator $ P_{\psi} (E) $.
 
 \item Divide the numerator by the denominator. 
 \end{enumerate} 
 It should be clear that none of these three 
 steps corresponds to an event that occurs in 
 the quantum system being studied. 
 Actually, none of these steps refers to 
 spacetime coordinates. 
 For example, it is makes no sense to ask how, 
 when or where the division in the third step 
 occurs in the quantum system. 
 Yet this is a valid algorithm for computing 
 the one conditional probability 
 $ P_{\psi} (F \,|\, E ) $. 
 Similar comments apply to the previous two 
 step algorithm, despite the fact that there 
 has been much discussion on the literature 
 about how, when and where ``collapse'' occurs. 
 Once more, quite simply it does not occur. 
 Only events occur.

 \vskip 0.4cm 
 \centerline{\bf Acknowledgment}
\vskip 0.2cm 
These thoughts are based on a talk which I gave at the BIRS Workshop 
 "Gravity, Noncommutative Geometry, Cosmology (23w5081)" held at Casa Matem\'atica Oaxaca
  (Oaxaca, Oax., M\'exico) on 
 August 22, 2023 entitled 
 {\em My Guesses about Quantum Gravity}.
I gratefully thank the organizers of 
this event 
for 
letting a complete non-expert speculate 
on such a notoriously difficult and problematical
topic.

\end{document}